\documentclass[10pt]{article}
\usepackage{epsfig}

\begin{document}

\title{Static exteriors for nonstatic braneworld stars} 
\author{J. Ponce de Leon\thanks{E-mail: jpdel@ltp.upr.clu.edu; jpdel1@hotmail.com}\\ Laboratory of Theoretical Physics, Department of Physics\\ 
University of Puerto Rico, P.O. Box 23343, San Juan, \\ PR 00931, USA} 
\date{Version 2, January  2008}

\maketitle

\begin{abstract}
We study possible static non-Schwarzschild exteriors for  nonstatic spherically symmetric stars in a Randall $\&$ Sundrum type II braneworld scenario. Thus, the vacuum region outside the surface of a star  is assumed to be a static solution to the equation $^{(4)}R = 0$, where $^{(4)}R $ is the scalar curvature of the $4$-dimensional Ricci tensor with spherical symmetry. 
Firstly, we show that for nonstatic spheres the standard matching conditions are much more restrictive    than for  static ones; they lead to a specific  requirement on the vacuum region outside of a  nonstatic star, that is absent in the case of static stars. Secondly, without making any assumption about the bulk, or the material medium inside the star, we prove the following theorem on the brane: for {\it any}  nonstatic spherical star, without rotation,  there are  only two possible static exteriors; these   are the Schwarzschild and the ``Reissner-Nordstr{\"o}m-like" exteriors. This is quite distinct from  the case of stars in hydrostatic equilibrium which admit a much larger family of non-Schwarzschild static exteriors.

\end{abstract}

\medskip



\newpage

\section{Introduction}

In recent years there has been an increased interest in theories that envision our world as embedded in a universe with more than four large dimensions \cite{Antoniadis}-\cite{JPdeLgr-qc/0105120v2}. The study of stellar structure and stellar evolution might constitute an important approach to predict observable effects from extra dimensions. 

An indispensable ingredient for these studies is to know how to describe the region outside of an isolated star. In general relativity this constitutes no problem because there is an unique spherically symmetric vacuum solution, namely the Schwarzschild exterior metric. 

However, in more that four dimensions, the effective equations for gravity in $4D$ are weaker than the Einstein equations in ordinary general relativity in the sense that they do not  constitute a closed set of differential equations. From a geometrical point of view this reflects the fact that there are many ways of producing, or embedding, a $4D$ spacetime in a given higher dimensional manifold, while satisfying the field equations \cite{JPdeLgr-qc/0512067}. From a physical point of view this is a consequence of nonlocal effects transported from the bulk to the brane by the projection of the $5D$ Weyl tensor onto the brane, which are unknown without specifying the properties of the metric in the bulk. 

As a consequence,  the  effective
picture in four dimensions allows the existence of different possible non-Schwarzschild scenarios for the description of the spacetime outside of a spherical star. In a recent paper we have studied various non-Schwarzschild exteriors in the context of static spherical stars \cite{arXiv:0711.0998v1[gr-qc]}. A number of interesting results emerged from that study. Among others, that the general relativistic upper bound on the gravitational potential $M/R < 4/9$, for perfect fluid stars, can significantly be increased in these exteriors. In particular, the upper bound is $M/R < 1/2$, $M/R < 2/3$ and $M/R < 1$ for the temporal Schwarzschild \cite{Cristiano}, \cite{Casadio}, spatial Schwarzschild \cite {Casadio} and  Reissner-Nordstr{\"o}m-like exteriors \cite{Cristiano}, respectively.

In this work, we concentrate our attention  on {\it nonstatic} spherical stars, without rotation,  in the context of the Randall $\&$ Sundrum type II braneworld scenario \cite{Randall2}. Our aim is to study possible static non-Schwarzschild exteriors.  This is crucial  in order to identify  the difference between stellar evolution and gravitational collapse in ordinary general relativity and braneworld models, which might shed some light on possible observational clues to detect effects from extra dimensions.

We will show here two specific features of nonstatic spheres in this scenario. Firstly, we show that for nonstatic spheres the standard matching conditions are much more restrictive    than for  static ones; they lead to a specific  requirement on the vacuum region outside of a  nonstatic star, that is absent in the case of static stars. Secondly, without making any assumption about the bulk, or the material medium inside the star, we prove the following theorem on the brane: for {\it any}  nonstatic spherical star, without rotation,  there are  only two possible static exteriors; these   are the Schwarzschild and the ``Reissner-Nordstr{\"o}m-like" exteriors. This is quite distinct from  the case of stars in hydrostatic equilibrium which admit a much larger family of non-Schwarzschild static exteriors. 

The paper is organized as follows. In section $2$ we review the effective field equations on the brane and the general static vacuum solutions. In section $3$ we present the stellar model. In section $4$ we discuss the boundary conditions and demonstrate the uniqueness of the Schwarzschild and Reissner-Nordstr{\"o}m-like exteriors for non-static spherical stars. In section $5$ we provide a simple example that illustrates the results. Section $6$ is a discussion and conclusions.   

\section{Field equations on the brane}

In order to make the paper self-consistent, let us restate some concepts that are essential in our discussion. This is also necessary, because some authors work with spacetime signature $(+, -, - , -)$, while others with $(-, +, +, +)$. Besides, there are different definitions for  the Riemann-Christoffel curvature tensor. As a consequence the Einstein field equations look different. For example, $G_{\mu\nu} = - (8\pi G/c^4) T_{\mu\nu}$  in \cite{Weinberg} and $G_{\mu\nu} = (8\pi G/c^4) T_{\mu\nu}$ in \cite{Landau}. 
In this  work  the spacetime signature is $(+, -, - , -)$; we follow the definitions of Landau and Lifshitz \cite{Landau}; and the speed of light $c$ is taken to be unity.

In the  Randall $\&$ Sundrum  braneworld scenario \cite{Randall2} the effective equations for gravity in $4D$ are obtained from dimensional reduction of the five-dimensional equations $^{(5)}G_{AB} = k_{(5)}^2 {^{(5)}T_{AB}}$. In this scenario our universe is identified with a singular hypersurface (the {\it brane}) embedded in a   $5$-dimensional anti-de Sitter bulk $(^{(5)}T_{AB} = - \Lambda_{(5)}g_{AB})$ with ${\bf Z}_{2}$ symmetry  with respect to the brane. The effective field equations in $4D$ are \cite{Shiromizu}
\begin{equation}
\label{EMT in brane theory}
^{(4)}G_{\mu\nu} = - {\Lambda}_{(4)}g_{\mu\nu} + 8\pi G T_{\mu\nu} +\epsilon k_{(5)}^4 \Pi_{\mu\nu} - \epsilon E_{\mu\nu},
\end{equation}
where $^{(4)}G_{\mu\nu}$ is the usual  Einstein tensor in $4D$; $\Lambda_{(4)}$ is the $4D$ cosmological constant, which is expressed in terms of the $5D$  cosmological constant $\Lambda_{(5)}$ and the brane tension $\lambda$, as 
\begin{equation}
\label{definition of lambda}
\Lambda_{(4)} = \frac{1}{2}k_{(5)}^2\left(\Lambda_{(5)} + \epsilon k_{(5)}^2\frac{ \lambda^2}{6}\right);
\end{equation}
$\epsilon$ is taken to be $- 1$ or $+ 1$, depending on whether the extra dimension is spacelike or timelike, respectively; $G$ is the Newtonian gravitational constant
\begin{equation}
\label{effective gravitational coupling}
8 \pi G = \epsilon k_{(5)}^4\frac{\lambda}{6};
\end{equation}
$T_{\mu\nu}$ is the energy momentum tensor (EMT) of matter confined in $4D$; $\Pi_{\mu\nu}$ is a tensor quadratic in $T_{\mu\nu}$ 
\begin{equation}
\label{quadratic corrections}
\Pi_{\mu\nu} = - \frac{1}{4} T_{\mu\alpha}T^{\alpha}_{\nu} + \frac{1}{12}T T_{\mu\nu} + \frac{1}{8}g_{\mu\nu}T_{\alpha\beta}T^{\alpha\beta} - \frac{1}{24}g_{\mu\nu}T^2;
\end{equation}
and $E_{\alpha \beta}$ is the projection onto the brane of the Weyl tensor in $5D$. Explicitly,  $E_{\alpha\beta} = {^{(5)}C}_{\alpha A \beta B}n^An^B$, where $n^{A}$ is the $5D$ unit vector $(n_{A}n^{A} = \epsilon)$ orthogonal to the brane. This quantity connects the physics in $4D$ with the geometry of the bulk.

Therefore, giving the EMT of matter in $4D$ is not enough to solve the above equations, because $E_{\alpha \beta}$ is unknown without specifying, {\it both} the metric in $5D$, and the way the $4D$ spacetime is identified \cite{JPdeLgr-qc/0512067}, \cite{JPdeLgr-qc/0511067}. In other words,  
the set of equations (\ref{EMT in brane theory}) is not closed in $4D$. The only quantity that can be specified without resorting to  the bulk metric, or the details of the embedding,  is the curvature scalar $^{(4)}R = {^{(4)}R}^{\alpha}_{\alpha}$,  because $E_{\mu\nu}$ is traceless. In particular, in empty space $(T_{\mu\nu} = 0, \Lambda_{(4)} = 0)$
\begin{equation}
\label{field eqs. for empty space}
^{(4)}R = 0.
\end{equation}
Thus, the braneworld theory provides  only one equation for the vacuum region outside the surface of a star. In the case of static spherically symmetric exteriors there are two metric functions, say  $g_{TT}$ and $g_{RR}$,  to be determined. As a consequence, (\ref{field eqs. for empty space}) admits a non denumerable infinity of solutions parameterized by some arbitrary function of the radial coordinate $R$ \cite{Viser}.  Since this is a second order differential equation for $g_{TT}$ and first order for $g_{RR}$, the simplest way for generating static solutions is to provide a smooth function of $R$ for $g_{TT}$. Then,  the field equation $^{(4)}R = 0$ reduces to a first order differential equation for $g_{RR}$, whose static solutions and their  general properties  have thoroughly been discussed in the literature \cite{Casadio}, \cite{Viser}, \cite{Dadhich}, \cite{Bronnikov}.

\section{The stellar model}
An observer in $4D$, who is confined to making physical measurements  in our ordinary spacetime, can interpret the effective  equations (\ref{EMT in brane theory})  as the conventional Einstein equations with an effective EMT, 
$T_{\mu\nu}^{eff}$, defined as  
\begin{equation}
\label{def. of effective EMT}
8 \pi GT_{\mu\nu}^{eff} \equiv - {\Lambda}_{(4)}g_{\mu\nu} + 8\pi G T_{\mu\nu} + \frac{48 \pi G}{\lambda} \Pi_{\mu\nu} - \epsilon E_{\mu\nu}.
\end{equation}
Thus, if we are dealing with a perfect fluid star with density $\rho$ and pressure $p$, then  the effective density and pressure are given by  $(\Lambda_{(4)} = 0)$
\begin{eqnarray}
\label{effective matter in terms of perfect fluid rho and p}
\rho^{eff} &=& \rho - \frac{\epsilon k_{(5)}^4}{48\pi G}\rho^2 - \frac{\epsilon E^{0}_{0}}{8\pi G}, \nonumber  \\
p^{eff}_{rad} &=& p - \frac{\epsilon k_{(5)}^4}{48 \pi G}(\rho + 2p)\rho + \frac{\epsilon  E_{1}^{1}}{8 \pi G}, \nonumber \\
p^{eff}_{\perp} &=& p - \frac{\epsilon k_{(5)}^4}{48 \pi G}(\rho + 2p)\rho + \frac{\epsilon  E_{2}^{2}}{8 \pi G}.
\end{eqnarray}
It should be noted that the effective matter quantities do not have to satisfy the regular energy conditions \cite{Bronnikov2}, because they involve terms of  geometric origin.

For a nonstatic, spherically symmetric distribution of matter in $4D$, the line element can be written as 
\begin{equation}
\label{general interior nonstatic metric}
ds^2 = e^{\nu(r, t)}dt^2 - e^{\lambda(r, t)}dr^2 - R^2(r, t)\left(d\theta^2 + \sin^2\theta d\phi^2\right).
\end{equation}
In a comoving frame, the field equations relate the effective density $\rho^{eff}$ and radial pressure $p_{rad}^{eff}$ to the mass function (from now on we set $G = 1$) 
\begin{equation}
\label{general mass function}
m(r, t) = \frac{R}{2}\left(1 + e^{- \nu}{\dot{R}}^2 - e^{- \lambda}R'^2\right),
\end{equation} 
as follows \cite{Misner}, \cite{Podurets}
\begin{equation}
\label{density from the mass function}
m' = 4\pi \rho^{eff}R^2 R',
\end{equation}
\begin{equation}
\label{pressure from the mass function}
\dot{m} = - 4\pi p_{rad}^{eff}R^2\dot{R},
\end{equation}
where dots and primes denote differentiation with respect to $t$ and $r$, respectively. Thus, 
\begin{equation}
\label{total mass interior to shell r}
m(r, t) = 4\pi \int_{0}^{r}{R^2 \rho^{eff}(\bar{r}, t)R' d\bar{r}},
\end{equation}
can be interpreted as the ``total mass-energy interior to shell $r$ at time $t$" measured by an observer riding in a given shell \cite{Gravitation}. 
We note that a similar expression, but for a static interiors, is used in \cite{Cristiano}.

We assume that the source is bounded, namely that 
the three-dimensional hypersurface $\Sigma$, defined by the equation
\begin{equation}
\Sigma: r - r_{b} = 0,
\end{equation}
where $r_{b}$ is a constant, separates the spacetime into two regions: the stellar interior described by (\ref{general interior nonstatic metric}) and an exterior vacuum region, which we assume is described by a static spherically symmetric line element in curvature coordinates. Namely,

\begin{equation}
\label{generic exterior}
ds^2 = A(R)dT^2 - B(R)dR^2 - R^2(d\theta^2 + \sin^2\theta d\phi^2),
\end{equation}
where the metric functions $A$ and $B$ are solutions of the field equation $^{(4)}R = 0$. 

In this coordinates the equation of the boundary takes the form
\begin{equation}
\label{equation of the boundary in external coordinates}
R = R_{b}(T),
\end{equation}
where $b$ stands for boundary. 
In the vacuum region, outside of the source the projection onto the brane of the Weyl tensor in $5D$ can be interpreted as an effective energy-momentum tensor $(\Lambda_{(4)} = 0)$, viz., 
\begin{equation}
\label{effective ETM outside the star}
T_{\mu \nu}^{eff} = - \frac{\epsilon}{8\pi} E_{\mu \nu}
\end{equation}  
where\footnote{In what follows, in order to simplify the notation, we will suppress the ``{\it eff}" over the matter quantities.}
\begin{equation}
\label{rho outside}
8\pi T_{0}^{0} = \frac{1}{R^2 B^2}\left[R\frac{dB}{dR} + B(B - 1)\right],
\end{equation}
\begin{equation}
\label{p radial outside}
8 \pi T_{1}^{1} = - \frac{1}{R^2 A B}\left[R \frac{dA}{dR} - A(B - 1)\right],
\end{equation}
\begin{equation}
\label{p tangential outside}
8\pi T_{2}^{2} =  8\pi T_{3}^{3} =  - \frac{1}{2R B}\left[\frac{1}{A}\frac{dA}{dR} - \frac{1}{B}\frac{dB}{dR} + \frac{R}{A}\frac{d^2 A}{dR^2} - \frac{R}{2A}\left(\frac{1}{A}\frac{dA}{dR} + \frac{1}{B}\frac{dB}{dR}\right)\frac{dA}{dR}\right].
\end{equation}
In addition, outside the surface
\begin{equation}
\label{Trace of EMT}
T_{0}^{0} + T_{1}^{1} + T_{2}^{2} + T_{3}^{3} = 0,
\end{equation}
which is a consequence of the fact that $E_{\mu\nu}$ is traceless. We are not going to discuss here the extension of these metrics to the bulk geometry. Finding an exact solution in $5D$ that is consistent with a particular induced metric in $4D$ is not an easy task. However, the existence of such a solution is guaranteed by Campbell-Maagard's embedding theorems \cite{SanjeevWesson}, \cite{Indefenceof}.

\section{Boundary conditions}
We recall that two regions of the spacetime are said to match across a separating non-singular surface $\Sigma$ if the first and second fundamental forms are continuous across $\Sigma$. These are essentially Israel's boundary  conditions in  vacuum\footnote{In the general case where the separating surface is a thin layer of matter, which is {\it not} the situation for stars, with surface energy-momentum tensor $S_{\alpha\beta}$ the extrinsic curvature is discontinuous across the layer.  If we denote the unit spacelike vector normal to $\Sigma$ by $n^{\mu}$, the induced metric on $\Sigma$  by $\lambda_{\alpha\beta} = (g_{\alpha \beta} - n_{\alpha} n_{\beta})$, and the  extrinsic curvature tensor on $\Sigma$ by $K_{\mu\nu} = \frac{1}{2}{\cal{L}}_{n} \lambda_{\mu\nu}$, then the discontinuity of $K_{\alpha\beta}$  is given by $({K_{\mu \nu}}_{|\Sigma^{+}} - {K_{\mu \nu}}_{|\Sigma^{-}}) = 8\pi (S_{\mu\nu} - \frac{1}{2}S \lambda_{\mu\nu})$. This condition and ${\lambda_{\mu\nu}}_{|\Sigma^{+}} = {\lambda_{\mu\nu}}_{|\Sigma^{-}}$
constitute the so-called Israel's boundary conditions \cite{Israel}.}. 

The continuity of the first fundamental form (the metric tensor induced on $\Sigma$) gives at once

\begin{equation}
\label{Rb as a function of t}
R_{b} = R(r_{b}, t) \equiv R_{b}(t),
\end{equation} 
and relates the coordinates $t$ and $T$, viz.,
\begin{equation}
\left(\frac{dT}{dt}\right)^2  = \frac{e^{\nu(r_{b}, t)}B(R_{b})}{A(R_{b})}\left(\frac{1}{B(R_{b})} + U_{b}^2\right),
\end{equation}
 where $U_{b} = e^{- \nu(r_{b}, t)/2}{\dot{R}_{b}}$.

Next, the continuity of the second fundamental form across $\Sigma$ (or the extrinsic curvature tensor on $\Sigma$) requires continuity of the mass function and the radial pressure. In curvature coordinates (\ref{generic exterior}) $r = R$ and $e^{\lambda} = B(R)$, therefore the mass function (\ref{general mass function}) reduces to 
\begin{equation}
\label{mass function for the exterior metric}
m(R) = \frac{R}{2}\left[1 - \frac{1}{B(R)}\right].
\end{equation}
The same can be obtained from (\ref{total mass interior to shell r}) after substituting (\ref{rho outside}) into it, and preforming the integration.
Thus, demanding the mass function to be continuous across the boundary we get
\begin{equation}
\label{continuity of the mass function}
m(r_{b}, t) = \frac{R_{b}(t)}{2}\left[1 - \frac{1}{B(R_{b}(t))}\right].
\end{equation}

Let us now evaluate the radial pressure at the surface. From (\ref{pressure from the mass function}) and (\ref{continuity of the mass function})  we find
\begin{equation}
\label{p = - T_1^1 in terms of B, the general case}
8\pi p_{rad}(r_{b}, t) = - \frac{1}{R^2 B^2}\left[R \frac{dB}{dR} + B(B - 1)\right],\;\;\;\;\mbox{evaluated at }\;\;\;R = R_{b}
\end{equation}
Thus, from (\ref{rho outside}) we find
\begin{equation}
\label{p = - T_1^1 in the general case}
p_{rad}(r_{b}) = - T^{0}_{0}(R)_{|R = R_{b}}.
\end{equation}
On the other hand, continuity of the second fundamental form requires $p_{rad} = - T_{1}^{1}$ at the boundary. 
Therefore, the exterior solution must satisfy
\begin{equation}
\label{second equation}
T_{0}^{0} = T_{1}^{1},\;\;\;\mbox{at}\;\;\;R = R_{b}.
\end{equation}

\subsection{The ``extra" requirement}

We should observe  that the relations (\ref{p = - T_1^1 in terms of B, the general case})-(\ref{second equation}), which are consequence of the effective field equation (\ref{pressure from the mass function}), are exclusive for nonstatic distributions of matter for the reason that    
in the case of static interiors there is no a similar relation between the mass function and pressure.
Therefore, for nonstatic distributions the boundary conditions impose the fulfillment of the extra requirement (\ref{second equation}), which is {\it absent} in the static case. 

The question is how to interpret this extra requirement. Firstly, we note that the hypersurface, (say $\Sigma_{(T_{0}^{0} = T_{1}^{1})}$) at which  $T_{0}^{0} = T_{1}^{1}$ has a   {\it fixed} radius, instead of being  a dynamical one as required by (\ref{equation of the boundary in external coordinates}), which is incompatible with the notion of nonstatic distribution of matter. Secondly, it is easy to demonstrate that $\Sigma_{(T_{0}^{0} = T_{1}^{1})}$
is either a horizon or a spherical surface of infinite radius. In order to show this, we use a technique employed by  Bronnikov {\it et al} \cite{Bronnikov}. Namely, we introduce a new radial coordinate $u$ defined by 
\begin{equation}
du = \sqrt{A(R)B(R)}\;dR,
\end{equation}
which leaves $\rho^{ext} = T_{0}^{0}$ and $p_{rad}^{ext} = - T_{1}^{1}$ invariant. The line element (\ref{generic exterior}) becomes
\begin{equation}
ds^2 = {\cal{A}}(u)dT^2 - \frac{du^2}{{\cal{A}}(u)} - R^2(u)\left[d\theta^2 + \sin^2\theta d\phi^2\right], 
\end{equation}
where ${\cal{A}}(u) = A(R)$ and $R(u) = R$. 
In this coordinates we find
\begin{equation}
T_{0}^{0} - T_{1}^{1} = - \frac{2 {\cal{A}}(u)}{R(u)}\left[\frac{d^2R(u)}{du^2}\right].
\end{equation}
This shows that the condition $T_{0}^{0} = T_{1}^{1}$, required by (\ref{second equation}),  is satisfied either (i) at a hypersurface $\Sigma_{(T_{0}^{0} = T_{1}^{1})}$ where  $A(R_{b}) = 0$, i.e., at a horizon, or (ii) at spatial infinity, $R = \infty$.

\medskip

From the above discussion, it is clear that (\ref{second equation}) is {\it not} a condition defining the boundary of a star. In what follows we will interpret it 
as an ``equation of state" for a static vacuum region outside of a {\it nonstatic} star. Namely,
\begin{equation}
\label{equation of state for the stellar exterior}
\rho^{ext} = - p_{rad}^{ext},\;\;\;\mbox{for}\;\;\;R \geq R_{b}(t), 
\end{equation}
where $\rho^{ext} = T_{0}^{0}$ and $p_{rad}^{ext} = - T_{1}^{1}$.  We remark that for static interiors this condition is gone.
 
\medskip

The preceding analysis is totally general. Therefore,  it is useful to  illustrate it with some examples.  Firstly, let us consider the  
``temporal Schwarzschild" metric \cite{Cristiano}, \cite{Casadio}
\begin{equation}
\label{temporal Schw exterior}
ds^2 = \left(1 - \frac{2{{M}}}{R}\right) dT^2 - \frac{(1 - 3{{M}}/2R)}{(1 - 2{{M}}/R)[1 - (3{{M}}/2R)\;c]}dR^2 - R^2 d\Omega^2,
\end{equation} 
where $c$ is an arbitrary dimensionless constant\footnote{In order to avoid misunderstanding, please note that $c$ has nothing to do with the velocity of light in vacuum.} and $M$ is the total gravitational mass measured by an observer at spatial infinity. For $c = 1$ it  reduces to the Schwarzschild vacuum solution of general relativity.

For this metric we find 
\begin{equation}
\rho^{ext} + p^{ext}_{rad} = \frac{3M(c - 1)(2M - R)}{4\pi R^4(2 - 3M/R)^2}.
\end{equation}
Thus, condition (\ref{second equation}) is satisfied everywhere for $c = 1$. Nevertheless, for $c \neq 1$ it is satisfied at the horizon  $R = 2M$  and at $R = \infty$, in agreement with the statement above.

Secondly, let us consider the ``spatial Schwarzschild"  metric \cite{Casadio} 
\begin{equation}
\label{spatialsSchw exterior}
ds^2 = \frac{1}{b^2}\left(b - 1 + \sqrt{1 - \frac{2 b M}{R}}\right)^2\;dT^2 - \left( 1 - \frac{2 b M}{R}\right)^{- 1}dR^2 - R^2 d\Omega^2,
\end{equation}
where $M$ is the total gravitational mass measured at spatial infinity and $b$ is a dimensionless constant. For $b = 1$, we recover the Schwarzschild exterior metric. 

For this metric we obtain
\begin{equation}
\rho^{ext} + p^{ext}_{rad} = \frac{b M(1 - b)}{4 \pi R^3 \left(b - 1 + \sqrt{1 - 2M b/R}\right)}.
\end{equation}
Thus, condition (\ref{second equation}) is satisfied everywhere for $b = 1$, but only at $R = \infty$ for $b \neq 1$, in accordance with the above discussion.

\subsection{Uniqueness of the Schwarzschild and Reissner-Nordstr{\"o}m-like exteriors for non-static spherical stars}  

We now proceed to show that, in the context of Randall-Sundrum's  single brane model scenario, there are only two possible static exteriors for a non-static spherical star, namely the Schwarzschild and the Reissner-Nordstr{\"o}m-like exteriors.

The requirement $T_{0}^{0} = T_{1}^{1}$ is equivalent to 
\begin{equation}
\label{A in terms of B}
A(R) = \frac{K}{B(R)}, 
\end{equation}
where $K$ is an arbitrary positive dimensionless constant. Now, from (\ref{Trace of EMT}) it follows that 
\begin{equation}
\label{equation for B in terms of T}
T_{0}^{0} =  - T_{2}^{2}. 
\end{equation}
Then, using (\ref{rho outside}) and (\ref{p tangential outside}), we get 
\begin{equation}
\frac{1}{B}\frac{d^2B}{dR^2} - \frac{2}{B^2}\left(\frac{dB}{dR}\right)^2 + \frac{4}{RB}\frac{dB}{dR} + \frac{2(B - 1)}{R^2} = 0 
 \end{equation}
The only solution to this equation is Reissner-Nordstr{\"o}m-like. Namely, 
\begin{equation}
\label{only solution for B}
B = \left(1 + \frac{C_{1}}{R} + \frac{C_{2}}{R^2}\right)^{- 1},
\end{equation}
where $C_{1}$ and $C_{2}$ are arbitrary constants of integration. What this means is that, according to the standard matching conditions, in the Randall $\&$ Sundrum II braneworld scenario the only possible static exteriors for  any  {\it nonstatic} spherical body, are the Schwarzschild and the Reissner-Nordstr{\"o}m-like exteriors.  We should emphasize the role of the field equation $^{(4)}R = 0$ (or $T = 0$) in obtaining this result.  

\section{Example}

In this section we present a simple model that illustrates the above calculations. With this aim,  
let us consider the case where effective density on the brane is spatially uniform, viz.,
\begin{equation}
\frac{\partial}{\partial r}\rho^{eff} = 0, \;\;\;\mbox{and}\;\;\;p_{rad}^{eff} = p_{\perp}^{eff}. 
\end{equation}
The most general line element corresponding to these assumptions, in comoving coordinates, is given by \cite{JPdeLUniformDensity}
\begin{equation}
\label{uniform density metric}
ds^2 = \frac{[r^2 + 2\dot{h}({t})/\dot{g}({t})]^2}{[\frac{1}{2}g({t})r^2 + h({t})]^2}\left\{C^2 d{t}^2 - \frac{1}{[r^2 + 2\dot{h}({t})/\dot{g}({t})]^2}\left(dr^2 + r^2 d\Omega^2\right)\right\}.
\end{equation}
where the functions $g(t)$ and $h(t)$, as well as the constant $C$,  are arbitrary except for the fact that they have to carry the  dimensions
\begin{equation}
[C] = L^{-2}, \;\;\;[g] = L^{-2}, \;\;\;[h] = L^{0},
\end{equation}
for  the metric coefficients to be dimensionless, as expected. For this metric we obtain
\begin{equation}
\label{mass function}
m(r, t) = R^3(r,t)\left(g(t)h(t) + \frac{{\dot{g}}^2(t)}{8C^2}\right), \;\;\;\mbox{with}\;\;\;R(r,t) = \frac{r}{\frac{1}{2}g(t)r^2 + h(t)}.
\end{equation}
Thus, at the boundary
\begin{equation}
R_{b}(t) \equiv R(r_{b},t) = \frac{r_{b}}{\frac{1}{2}g(t)r_{b}^2 + h(t)}, 
\end{equation}
and
\begin{equation}
\label{m in terms of g and h}
1 - 2R_{b}^2\left[g(t)h(t) + \frac{{\dot{g}}^2(t)}{8C^2}\right] = \frac{1}{B(R_{b})}.
\end{equation} 
The function $g(t)$ can be expressed in terms of $R_{b}$, 
\begin{equation}
\label{g in terms of R and h}
g(t) = - \frac{2\left[R_{b}(t)h(t) - r_{b}\right]}{r_{b}^2R_{b}(t)}.
\end{equation}
Without loss of generality, in order to simplify the equations bellow, instead of $h(t)$ it is preferable to work with  the function $\beta(t)$ defined as
\begin{equation}
\label{definition of beta}
\beta(t) = \frac{\alpha}{2}g(t) - h(t),
\end{equation}
where $\alpha$ is a parameter with the appropriate units. Substituting these expressions in (\ref{m in terms of g and h}) we obtain the equation that governs the evolution of the boundary, viz., 

\begin{equation}
\label{dR/dt in the general case with B}
\left(\frac{dR_{b}}{dt}\right)^2 = \frac{1}{[r_{b} + R_{b}^2(d\beta/dR_{b})]^2}\left[{R_{b}^2C^2(\alpha - r_{b}^2)^2} + {4R_{b}^3C^2r_{b}\beta(\alpha - r_{b}^2)} + 4R_{b}^4C^2r_{b}^2\beta^2 - \frac{R_{b}^2C^2(\alpha + r_{b}^2)^2}{B(R_{b})}\right],
\end{equation}
where $d\beta /dR_{b} =  (d\beta/dt)/{\dot{R}}_{b}$.

Substituting (\ref{m in terms of g and h}) into (\ref{density from the mass function}) and (\ref{pressure from the mass function}) we find the energy density and pressure inside the source as follows

\begin{equation}
\label{density in the general case with B}
\rho = \frac{3(B - 1)}{8\pi R_{b}^2 B},
\end{equation}

\begin{equation}
\label{New pressure in general with B}
p =  \frac{\left[(r_{b}^2 - r^2)\beta R_{b} + r_{b}(\alpha + r^2)\right]R_{b}(dB/dR_{b}) - B(B - 1)\left[3 (r_{b}^2 - r^2)R_{b}^2(d\beta/dR_{b}) +2(r_{b}^2 - r^2)\beta R_{b} - r_{b}(\alpha + r^2)\right]}{8\pi\left[(r_{b}^2 - r^2)R_{b}^2(d\beta/dR_{b}) - (\alpha + r^2)r_{b}\right]R_{b}^2B^2},
\end{equation}
where $B$ is evaluated at the boundary, i.e., $B = B(R_{b})$. For example, in general relativity where the vacuum region outside the surface is the  Schwarzschild metric, we set $B(R_{b}) = (1 - 2M/R_{b})^{- 1}$, and obtain
\begin{equation}
\label{density for Schw exterior}
\rho_{Schw} = \frac{3M}{4\pi R_{b}^3},
\end{equation}
which is a well known expression.
For the pressure we get
\begin{equation}
\label{pressure for the Schw exterior}
p_{Schw} = \frac{3M\left[\beta + R_{b}(d\beta/dR_{b})\right] (r_{b}^2 - r^2)}{4\pi \left[(\alpha + r^2)r_{b} - (r_{b}^2 - r^2)R_{b}^2(d\beta/dR_{b})\right]R_{b}^2},
\end{equation}
where $R_{b}$ is a solution of
\begin{equation}
\label{Schw exterior}
\left(\frac{dR_{b}}{dt}\right)^{2}_{|Schw} =  \frac{{2MC^2(\alpha + r_{b}^2)^2}R_{b} - 4\alpha r_{b}^2C^2 R_{b}^2 + {4\beta C^2 r_{b} (\alpha - r_{b}^2)}R_{b}^3 + 4\beta^2r_{b}^2C^2 R_{b}^4}{\left[r_{b}  + R_{b}^2(d\beta/dR_{b})\right]^2}.
\end{equation}
Coming back to our problem, evaluating the pressure (\ref{New pressure in general with B})
at the boundary $r = r_{b}$ we obtain
\begin{equation}
\label{p at the boundary}
8\pi p(r_{b}) = - \frac{R_{b}(dB/dR_{b}) + B^2 - B}{R_{b}^2 B^2} = - T^{0}_{0}(R)_{|R = R_{b}},
\end{equation}
which is exactly what we obtained in the general case (\ref{p = - T_1^1 in terms of B, the general case}), (\ref{p = - T_1^1 in the general case}). Similar results can be obtained in models with non-uniform effective density. 

\section{Discussion and conclusions}

The continuity of the second fundamental form and the field equation (\ref{pressure from the mass function}) require $T_{0}^{0} = T_{1}^{1}  $ at the surface  of a nonstatic star (\ref{second equation}). However, this equation cannot be considered as a condition defining the boundary, because it can only be satisfied either at a horizon or at spatial infinity. This was shown in section $3.1$.

Consequently, in order to be able to match a nonstatic interior with a static exterior we {\it have} to assume that (\ref{second equation}) is satisfied {\it everywhere}, not only at spatial infinity or at a horizon. 
This  constitutes an independent equation outside the surface of a nonstatic star, namely  $\rho^{ext} + p_{rad}^{ext} = 0$, which in addition to (\ref{field eqs. for empty space}) provides a complete set of equations to determine the two metric functions, A(R) and B(R) in our notation. We have found that the  only static solution to these equations is a Reissner-Nordstr{\"o}m-like metric, which is  given by  (\ref{A in terms of B}) and (\ref{only solution for B}). 

It should be noted that for stars in equilibrium, the space of non-Schwarzschild static exteriors allowed by the boundary conditions is much more general than in the nonstatic case \cite{arXiv:0711.0998v1[gr-qc]}. The natural question to ask here is, why?. If the boundary conditions are expressed in terms of the continuity of the first and second fundamental forms, why do we get ``different" results for static and nonstatic stars?

The answer to this question is found not in the boundary conditions, but in the field equations: in the nonstatic case there is a specific relation between the  time derivative of the mass function and the radial pressure, which is given by (\ref{pressure from the mass function}). In the static case there is no such relation. Therefore, in the nonstatic case there is an additional expression to be satisfied, namely (\ref{second equation}), which is absent in the static one.    
 As a result the space of solutions of the boundary conditions for stars in equilibrium is much greater than the one for nonstatic stars. 

From a physical point of view this is a consequence of the interconnection between the brane and the bulk. Indeed,  Israel's boundary conditions and the ${\bf Z}_{2}$ symmetry applied to the brane relate $T_{\mu\nu}$, the EMT of the fields in $4D$, with the extrinsic curvature $K_{AB} = \frac{1}{2}\partial g_{AB}/\partial y$ of the brane, where $y$ denotes the coordinate along the extra dimension. Thus, if $T_{\mu\nu}$ varies with time, one would expect the metric in the bulk $g_{AB}$  as well as $E_{\mu\nu}$, which carries non-local gravitational effects from the bulk to the brane, to be, in general, nonstatic. But $E_{\mu\nu}$ is the effective ETM in the exterior region (\ref{effective ETM outside the star}). Therefore,  in general, the exterior spacetime around a nonstatic spherical star is, is expected to be nonstatic as well. What is amazing here is that, despite of this chain of interaction between the bulk and the brain, one can still find some static exteriors for nonstatic stars.

Our results suggest that the temporal and spatial Schwarzschild metrics, as well as other possible static exteriors, are limiting configurations (in time) of non-static exteriors. In other words, if the contraction of a star comes to a halt and it reaches hydrostatic equilibrium, one would expect that a nonstatic exterior will tend to  one of the possible static exteriors.


\begin{thebibliography}{99}
\bibitem{Antoniadis}{I. Antoniadis, {\em Phys. Lett.} {\bf B246}, 3171(1990).}
\bibitem{Maartens1}{R. Maartens, {\em Phys. Rev.} {\bf D62}, 084023 (2000); hep-th/0004166.}
\bibitem{Maartens2}{Roy Maartens, Frames and Gravitomagnetism, ed. J Pascual-Sanchez et al. (World Sci., 2001), p93-119; gr-qc/0101059.}




\bibitem{Dadhich1}{Naresh Dadhich and S.G. Gosh, {\em Phys. Lett.} {\bf B518}, 1(2001); hep-th/0101019.}
\bibitem{Govender}{M. Govender and N. Dadhich, {\em Phys.Lett.} {\bf B538}, 233(2002);  hep-th/0109086.}
\bibitem{Cristiano}{C. Germani and Roy Maartens, {\em Phys. Rev.} {\bf D64},  124010(2001); 
hep-th/0107011.}
\bibitem{Bruni}{M. Bruni, C. Germani and R. Maartens, {\em Phys. Rev. Lett.}
{\bf 87}, 231302(2001);   gr-qc/0108013.}

\bibitem{Kofinas}{G. Kofinas and E. Papantonopoulos, {\em J. Cosmol. Astropart. Phys.} {\bf 12},  11(2004); gr-qc/0401047.}

\bibitem{Wesson 1}{P.S. Wesson, {\em G. Rel. Gravit.} {\bf 16}, 193(1984).}
\bibitem{JPdeL 1}{J. Ponce de Leon, {\em Gen. Rel. Grav.} {\bf 20}, 539(1988).}
\bibitem{Wesson and JPdeL}{P.S. Wesson and J. Ponce de Leon, {\em J. Math. Phys.} {\bf 33}, 3883(1992).}
\bibitem{Coley1}{A.A. Coley and D.J. McManus, {\em J. Math. Phys.} {\bf 36}, 335(1995).}
\bibitem{Overduin}{J.M. Overduin and P.S. Wesson, {\em Phys. Reports} {\bf 283}, 303(1997).}
\bibitem{Coley2}{A.P. Billiard and A.A. Coley, {\em Mod. Phys. Lett.} {\bf A12}, 2121(1997).}
\bibitem{Wesson book}{P.S. Wesson, {\em Space-Time-Matter} (World Scientific Publishing Co. Pte. Ltd. 1999).}
\bibitem{JPdeLgr-qc/0105120v2}{J. Ponce de Leon, {\em Int.J.Mod.Phys.} {D11},  1355(2002);  gr-qc/0105120.}

\bibitem{JPdeLgr-qc/0512067}{J. Ponce de Leon, {\em Class.Quant.Grav.} {\bf 23},  3043(2006); gr-qc/0512067.}
\bibitem{arXiv:0711.0998v1[gr-qc]}{J. Ponce de Leon, ``Stellar models with Schwarzschild and non-Schwarzschild vacuum exteriors" To be published in {\em Gravitation $\&$ Cosmology}; 	arXiv:0711.0998v1 [gr-qc].}
\bibitem{Casadio}{R. Casadio, A. Fabbri and L. Mazzacurati, {\em Phys.Rev.} {\bf D65}, 084040(2002);  
gr-qc/0111072.}
\bibitem{Randall2}{L. Randall and R. Sundrum, {\em Phys. Rev. Lett. } {\bf 83}, 4690(1999); hep-th/9906064.}
\bibitem{Weinberg}{Steven Weinberg, {\em Gravitation and Cosmology} (John Wiley and Sons, Inc. 1972).}
\bibitem{Landau}{L.D. Landau and E.M. Lifshitz, {\em The Classical Theory of Fields}, Fourth Edition (Butterworth-Heinemann, 2002).}
\bibitem{Shiromizu}{T. Shiromizu, Kei-ichi Maeda and Misao Sasaki, {\em Phys. Rev.} {\bf D62}, 02412(2000); gr-qc/9910076.}
\bibitem{JPdeLgr-qc/0511067}{J. Ponce de Leon, {\em Mod. Phys. Lett.} {\bf A21}, 947(2006); gr-qc/0511067.}
\bibitem{Viser}{M. Visser and D. L. Wiltshire, {\em Phys.Rev.} {\bf D67}, 104004(2003); hep-th/0212333.}
\bibitem{Dadhich}{N. Dadhich, R. Maartens, P. Papadopoulos and V. Rezania, {\em Phys.Lett.} {\bf B487},  1(2000); 
hep-th/0003061.}
\bibitem{Bronnikov}{K.A. Bronnikov, H. Dehnen and V.N. Melnikov, {\em Phys.Rev.} {\bf D68},   024025(2003); gr-qc/0304068.}
\bibitem{Bronnikov2}{K.A. Bronnikov and S-W Kim, {\em Phys.Rev.} {\bf D67},  064027(2003); gr-qc/0212112.}

\bibitem{Misner}{C.W. Misner and D.H. Sharp, {\em Phys. Rev.} {\bf 136}, B571(1964).}
\bibitem{Podurets}{M.A. Podurets, {\em Astron. Zh.} {\bf 41}, 28(1964); {\em Soviet Astron} {\bf 8}, 19(1964).}
\bibitem{Gravitation}{C.W Misner, K.S. Thorne and J.A. Wheeler, {\em Gravitation}, page 858 (W.H. Freeman and Company, 1973). }


\bibitem{Israel}{W. Israel, {\em Nuovo Cim.} {\bf B44}, 1(1966);[Erratum-ibid. {\bf B48}, 463(1967)].}



\bibitem{SanjeevWesson}{S.S. Seahra and P.S. Wesson, {\em Class.Quant.Grav.} {\bf 20} 1321(2003); gr-qc/0302015.}
\bibitem{Indefenceof}{P.S. Wesson, ``In Defense of Campbell's Theorem as a Frame for New Physics"; gr-qc/0507107.}
\bibitem{JPdeLUniformDensity}{J. Ponce de Leon, {J. Math. Phys.} {\bf 27}, 271(1986).}


\end{thebibliography}
\end{document}